# Magnetic and Magnetocaloric Properties of a $C_{20}$ Fullerene Structure: Monte Carlo Study

A. Jabar [1,2], S. Benyoussef [3], and L. Bahmad [3,*]

[1]Laboratory of Condensed Matter and Interdisciplinary Sciences (LaMCScI), Faculty of Sciences, Mohammed V University, Av. Ibn Batouta, B. P. 1014 Rabat, Morocco

[2]LPHE-MS, Science Faculty, Mohammed V University in Rabat, Morocco

[3]LPMAT, Faculty of Sciences Aïn Chock, Hassan II University of Casablanca, B.P. 5366 Casablanca, Morocco

*Corresponding author: l.bahmad@um5r.ac.ma (L.B.)

**Abstract**

One of the most active classes of nanostructures is Fullerene $C_{20}$, which has been exploited as an active component in significant applications. In this investigation, we used Monte Carlo simulations to investigate the magnetic and magnetocaloric properties of the mixed spins 2 and 3/2 Fullerene $C_{20}$ system. Ferrimagnetic and ferromagnetic phases are stable, according to the ground state phase diagrams that have been constructed. The behavior of the magnetizations and the derivative of magnetization, in particular, have shown the impact of rising temperature. Additionally, we found an increase of the reduced Curie temperature to reach $t_C \approx 3$, when the interactions between the spins S are increased. For numerous reduced external magnetic fields and reduced temperatures, the magnetic entropy variations are studied. The Relative Cooling Power (RCP) is calculated. It is demonstrated that the reduced exchange coupling interactions p and r lead to an increase in the lowered magnetic coercive field.

1. **Introduction**

Nanotechnology carbon-based materials are a prominent area of research, with the Fullerene family holding one of the top positions [1]. Applications for Fullerenes can be found in the nanoscale storage of ions and gases. On the other hand, they can also be utilized in lithium-ion batteries as cathode/anode electrodes [2]. Additionally, according to ref. [3], substitutional impurities are the most practical way to improve the required electrical characteristics of the **$C_{20}$** cage for use in nano-electronics and solar cell applications. Since rare gases are only weakly adsorbed on pure **$C_{20}$**, theoretical findings have demonstrated that this nano-cage cannot serve as an effective sensor for sensing and detecting rare gases.

In earlier research, caged Fullerene **$C_{20}$** solids were effectively synthesized using the ion beam irradiation approach. These findings suggest that **$C_{20}$** molecules can function as carbon cages for dodecahedrons in condensed phases, particularly when functioning as the building blocks of hexagonal crystals [4].

Additionally, the microstructures and sequential growth characteristics of tiny carbon nanocrystal particle (**$C_{20}$, $C_{24}$, $C_{26}$**) series, which were produced in an ultrahigh molecular weight polyethylene target by $C^{2+}$ ion beam irradiation, were investigated [5]. Furthermore, it was demonstrated in reference that fullerenes are graphitic cage structures with precisely twelve pentagons. Thus, **$C_{20}$** is the smallest fullerene that is made entirely of pentagons.

In ref. [6], it was demonstrated that the per hydrogenated form of the Fullerene (dodecahedrane $C_{20}H_{20}$) may be converted into the cage-structured Fullerene **$C_{20}$** by swapping the hydrogen atoms for very weakly attached bromine atoms and then performing gas-phase debromination. Several complimentary analysis methods, including high-resolution transmission electron microscopy, electron diffraction, laser desorption post-ionization, and Raman spectroscopy, have been used to characterize the structures of the **$C_{20}$** crystal. All of the hetero-fullerenes are actual minima, as evidenced by the structural stabilities, geometry, and electronic characteristics of $C_{20}$ as determined by density functional theory (DFT) computations [7].

On the other hand, the Monte Carlo simulations have been used to deduce the magnetic properties of some Fullerene-Like Structures: X20, X60, or X70 [8], in different fullerenes Xn nano-structures: Monte Carlo study [9]. Also, the phase diagrams and magnetic properties of a double fullerene structure with core/shell have been investigated in Ref. [10]. Moreover, the magnetic properties of a bi-fullerene-like structure, X60-Y60, with RKKY interactions in the Blume-Capel model have been the subject of Ref. [11]. In addition, the mixed spins in a Fullerene X30_Y30-Like structure, have been illustrated in Ref. [12].

It has been demonstrated that fullerenes $C_{20}$ with the $\bar{5}\bar{3}m$ symmetry can continuously deform into seven different low-symmetry species of the original "mother" fullerene molecule [13].

The geometrical arrangement of silicon impurities within the $C_{20}$ cage has been found to affect the electronic characteristics of the material just as much as silicon impurity concentrations. Additionally, by modifying the energy gap, the tuning of the electronic characteristics alters the charge transport and the absorption spectra for the $C_{20}$ cage significantly. On the other hand, the structure and electronic characteristics of TM@$C_{20}$ (TM = Sc, Ti, V, Cr, Mn, Fe, Co, Ni, Cu, and Zn) complexes were studied using DFT, and the calculated vibrational frequencies for the Ti, V, and Mn. These studies demonstrate the stability of the complexes and the ability to stabilize $C_{20}$ fullerene with $I_h$ symmetry.

According to Ref. [10] on high-level coupled-cluster calculations for the anion states of the smallest fullerene $C_{20}$, there are five bound electronic states for the $C_{20}$ anion at the equilibrium configuration of the $C_{20}$ neutral ground-state $D_{3d}$. Doping $C_{20}$ fullerene with hydrogen allows for effective adjustment of the electronic distribution and border orbitals in the original $C_{20}$ [14].

The magnetic moment of titanium-doped **C₂₀** is identical to that of titanium in isolation. The Co-doped **C₂₀** fullerene exhibits the strongest hybridization of any doped cage. Comparatively speaking to other clusters, the iron-doped **C₂₀** has extremely strong electrical and magnetic dipoles [15]. Depending on the type of connection between fullerene and carbon nanotubes, the total magnetic moment of nano-buds ranges from 0.28 µB to 4.00 µB [16]. The geometry $X_{20}$, $X_{60}$, and $X_{70}$, where the symbol X can be assigned to any magnetic atom, has been explored for its magnetic properties using a Monte Carlo simulation. It was discovered that when the number of atoms in the system under study increases, the critical temperature and the coercive field also rise [17]. Several existing literature used Monte Carlo simulations (MCS) to examine the mixed Ising model in order to theoretically analyze the magnetic properties, magnetocaloric properties, and phase transitions of different magnetic systems [18-25].

The major goal of our current research is to use theoretical calculations at the magnetic and magnetocaloric levels to look into the unexplored facets of the physical properties of **C₂₀**. The following is how the paper is set up: In section 2, we describe the details of the computational method. In section 3, we examine the obtained results of the magnetic and magnetocaloric properties of the studied materials. Section 4 is devoted to conclusions.

## 2. Ising model

The Fullerene **C₂₀** structure with mixed spin S and σ, is illustrated in Fig.1. The corresponding geometry consists of exactly 12 pentagons. The Hamiltonian of this system has first nearest neighbor interactions $J_{S\sigma}$, $J_{\sigma\sigma}$ and $J_{SS}$, and the crystal field $\Delta$ and external magnetic field h. Such Hamiltonian is given by:

$$H = -J_{S\sigma}\sum_{<i,j>} S_i\sigma_j - J_{\sigma\sigma}\sum_{<j,k>}\sigma_j\sigma_k - J_{SS}\sum_{<i,l>}S_iS_l - \Delta\left(\sum_i S_i^2 + \sum_i \sigma_i^2\right) - h\left(\sum_i S_i + \sum_j \sigma_j\right) \quad (1)$$

Where the notation <i, j> stands for the first nearest neighbor between site i and j. the values of moments spin S are ±2,0 and ±1 for σ are ±3/2 and ±1/2. In the full text, the parameter $J_{σσ}$ is taken 1.0 and the new reduced parameters are: p=$J_{Sσ}$/$J_{σσ}$, r=$J_{SS}$/$J_{σσ}$, d=Δ/$J_{σσ}$, t=T/$J_{σσ}$.

We apply Monte Carlo simulations (MCS) based on the Metropolis algorithm. In addition, free conditions on the lattice have been applied to the structure illustrated in Fig. 1.

First, the internal energy per site

$$E = \frac{\langle H \rangle}{N} \tag{2}$$

The partials and total magnetizations of the Fullerene **C20** are,

$$M_S = \left\langle \frac{1}{N_S} \sum_i S_i \right\rangle \tag{3}$$

$$M_σ = \left\langle \frac{1}{N_σ} \sum_i σ_i \right\rangle \tag{4}$$

$$M = \frac{N_S \times M_S + N_σ \times M_σ}{N_S + N_σ} \tag{5}$$

The partials and total magnetic susceptibilities of the Fullerene **C20** are,

$$\mathcal{X}_S = \beta(<M_S^2> - <M_S>^2) \tag{6}$$

$$\mathcal{X}_σ = \beta(<M_σ^2> - <M_σ>^2) \tag{7}$$

$$\chi = \frac{N_S \mathcal{X}_S + N_σ \mathcal{X}_σ}{N_S + N_σ} \tag{8}$$

The magnetic heat capacity of the Fullerene **C20** is given by:

$$C_m = \beta^2 \left( \langle E^2 \rangle - \langle E \rangle^2 \right) \tag{9}$$

Magnetocaloric effect can be related to the magnetic properties of the material through a thermodynamic Maxwell relation:

$$\left(\frac{\partial S}{\partial h}\right)_T = \left(\frac{\partial M}{\partial T}\right)_h \tag{10}$$

The magnetic entropy of a material can be calculated from this equation:

$$S(T,h) = \int_0^T \frac{C_m}{T} dT \tag{11}$$

The magnetic entropy changes between h different to zero and h = 0 is:

$$\Delta S_m = S_m(T,h) - S_m(T,0) = \int_0^h \left(\frac{\partial M}{\partial T}\right)_{h_i} dh \tag{12}$$

where $\beta = \dfrac{1}{k_B T}$, T is the absolute temperature and $k_B = 1$ is the Boltzmann's constant.

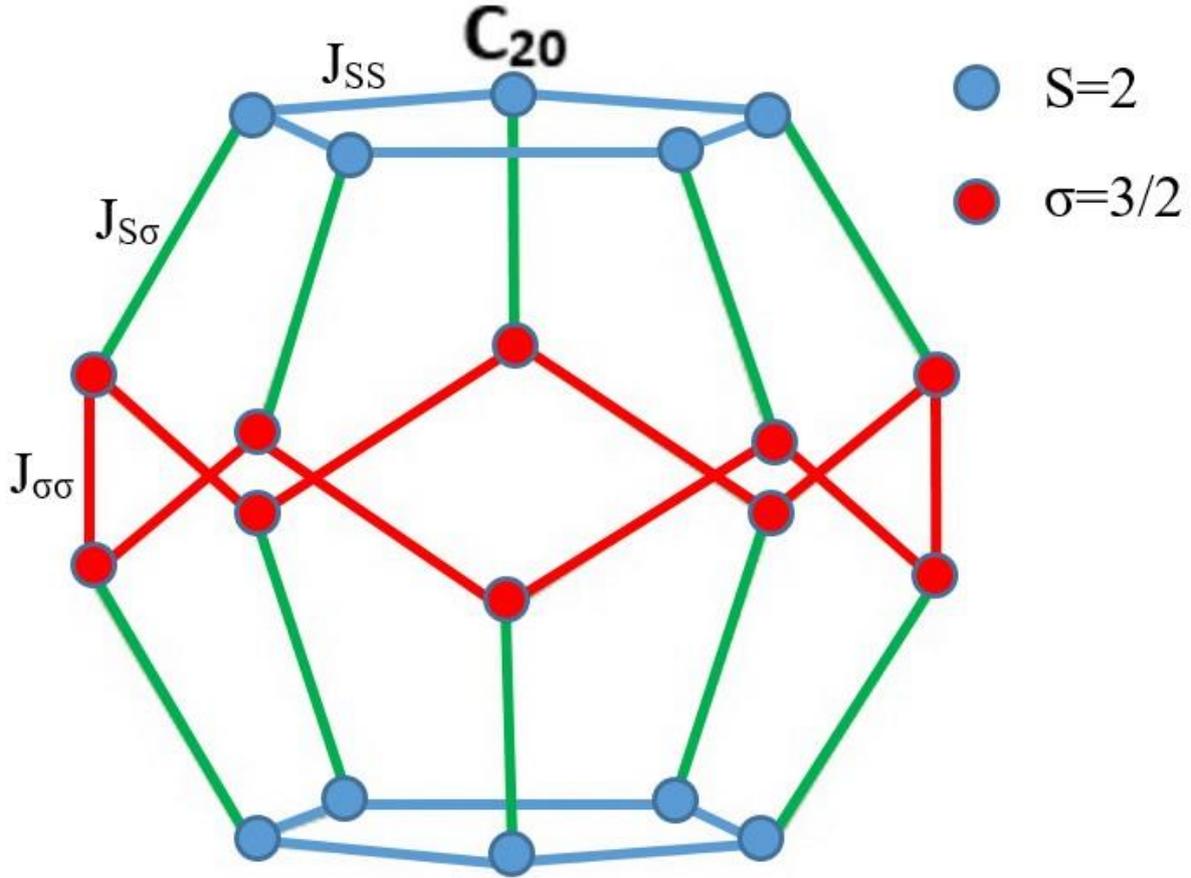

***Fig.1****: $C_{20}$ system with mixed spin S and σ.*

### 3. Results and discussions

The ground states of the Fullerene **$C_{20}$** system described by the Hamiltonian (1) are obtained by the computation and comparison of energies of all possible phases of (S, σ). The results of calculations reveal that there are numerous stable phases presented in Figs. 2(a), 2(b) and 2(c). In Fig. 2(a), we show stability regions for each (S, σ) phase in the (r, p) plane, corresponding to the phase diagram for fixed parameters for $h/J_{\sigma\sigma}=0$, d=0. We found that (±2, ±3/2), (±2, ∓3/2), (±1, ±3/2) and (±1, ∓3/2) exhibited ferri-magnetic stable phases and the phases (0, ±3/2) are ferro-magnetic stable ones. For r < 0, three phase transitions can be found (from ((±2, ∓3/2) to (±1, ∓3/2)), (from ((±1, ∓3/2) to (0, ±3/2)), (from ((0, ±3/2) to (±1, ±3/2)) and (from ((±1, ±3/2) to (±2, ±3/2)), when p increases. For r > 0, only one phase transition is possible (from ((±2, ∓3/2) to (±2, ±3/2)). In the second part, we show stability regions for each (S, σ) phase in the (r, h/J$_{\sigma\sigma}$) plane. We plot in Fig.2 (b) the phase diagram for fixed parameters for p=1.0, d=0. We deduce that the configurations (-2, -3/2), (-2, 3/2), (2, -3/2), (2, 3/2), (1, 3/2), (1, 3/2) and ($\pm$1, $\mp$3/2) are a ferri-magnetically stable and that (0, $\pm$3/2) are a ferromagnetic stable phases. In Fig. 2(c), the magnetic phase diagrams are obtained in plane (d, h/J$_{\sigma\sigma}$). We found fifteen ferrimagnetic stable phases named : (-2,-3/2), (-1,-3/2), (-2,3/2), (-2, $\pm$1/2), (-1,1/2), (-1,-1/2), (1,-3/2), (-1,3/2), (2,-3/2), (1,-1/2), (1,1/2), (2, $\pm$1/2), (1,3/2) and (2,3/2). Also, we found four ferromagnetic stable phases named: (0,-3/2), (0,3/2), (0,-1/2) and (0,1/2). Each configuration corresponds to a different region, and a complete symmetry is discovered, see Fig.2(c) for a fixed values r=1.0 and p=1.0.

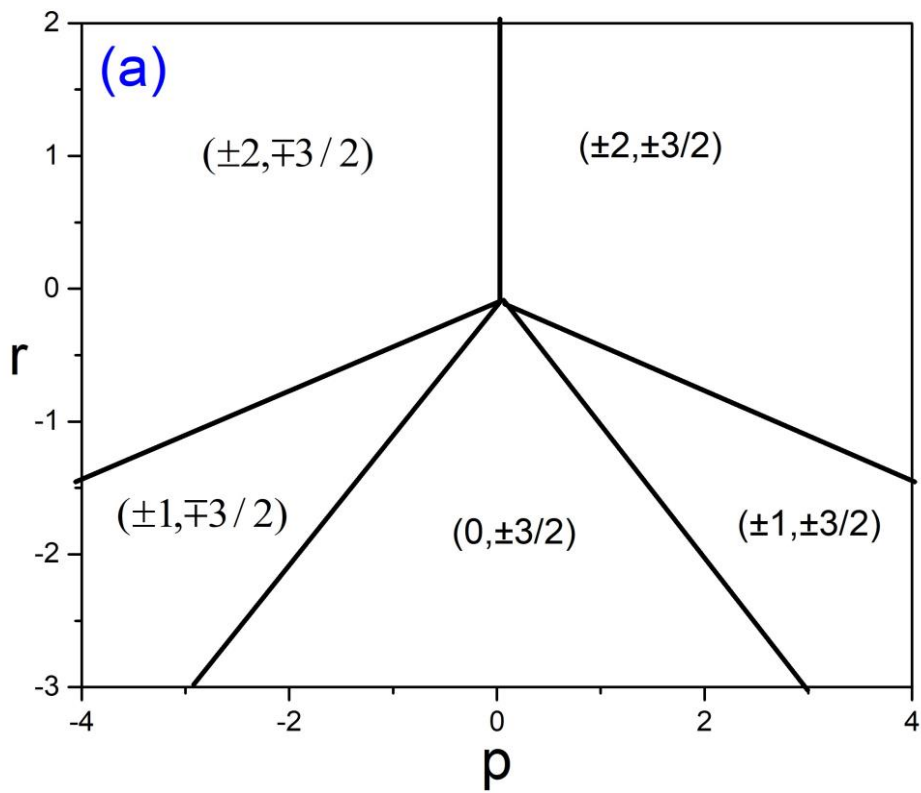

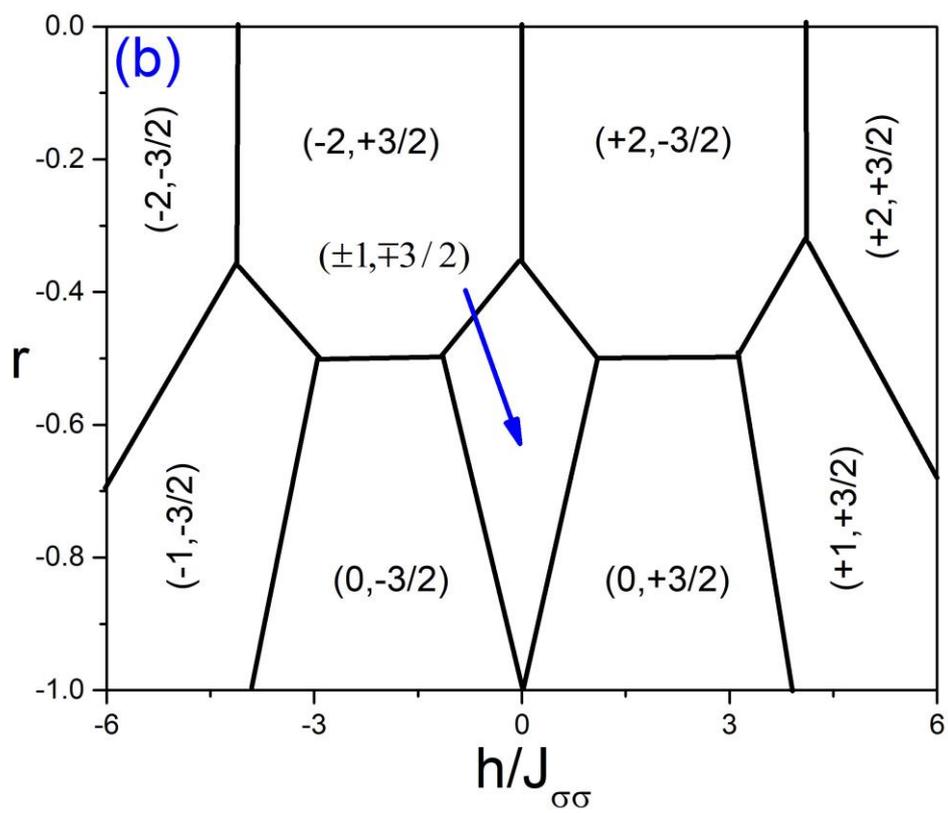

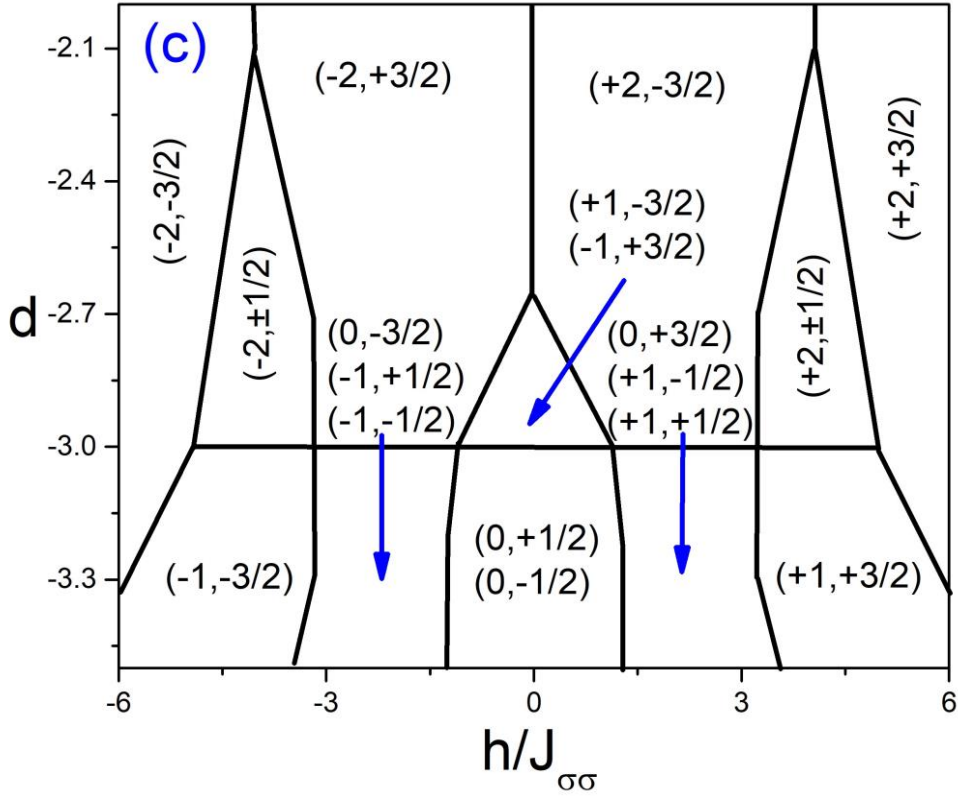

*Fig. 2*: Ground state phase diagrams of the $C_{20}$ system. (a) in the plane $(r, p)$ for $d=0.0$ and $h/J_{\sigma\sigma}=0.0$ (b) in the plane $(r, h/J_{\sigma\sigma})$ for $p=1.0$ and $d=0.0$ (c) in the plane $(d, h/J_{\sigma\sigma})$ for $r=1.0$ and $p=1.0$.

We performed the Monte Carlo simulations using the structure of Fig. 1, consisting on free conditions. Each site containing either S or σ spins, was visited $10^6$ times corresponding to the Monte Carlo steps MCS. Each step produced a new configuration of the system according to the Boltzmann distribution. We averaged our physical parameters over the last MCS when discarding the first $10^5$ MCSs.

In Fig. 3(a), the total magnetizations of the Fullerene $C_{20}$ system as a function of the reduced temperature is determined under different reduced external magnetic field $h/J_{\sigma\sigma}$ from 0.1 to 1.0 and for $r=0.2$, $p=1.0$ and $d=0$. Furthermore, once the reduced temperature increase, the curves illustrate a rapid decrease observed around a continuous phase transition where only the second phase transition is shown.

Accordingly, the derivative of magnetization (dM/dT) versus reduced temperature curves, plotted in Fig. 3(b), show that the reduced Curie Temperature $t_C$ increases when increasing the $h/J_{\sigma\sigma}$ from 0.1 to 1.0. Furthermore, the reduced Curie Temperature is very sensitive to the reduced external magnetic field and its value is as following: $t_C(h/J_{\sigma\sigma}=0.1) < t_C(h/J_{\sigma\sigma}=0.2) < t_C(h/J_{\sigma\sigma}=0.3) < t_C(h/J_{\sigma\sigma}=0.4) < t_C(h/J_{\sigma\sigma}=0.5) < t_C(h/J_{\sigma\sigma}=0.6) < t_C(h/J_{\sigma\sigma}=0.7) < t_C(h/J_{\sigma\sigma}=0.8) < t_C(h/J_{\sigma\sigma}=0.9) < t_C(h/J_{\sigma\sigma}=1)$.

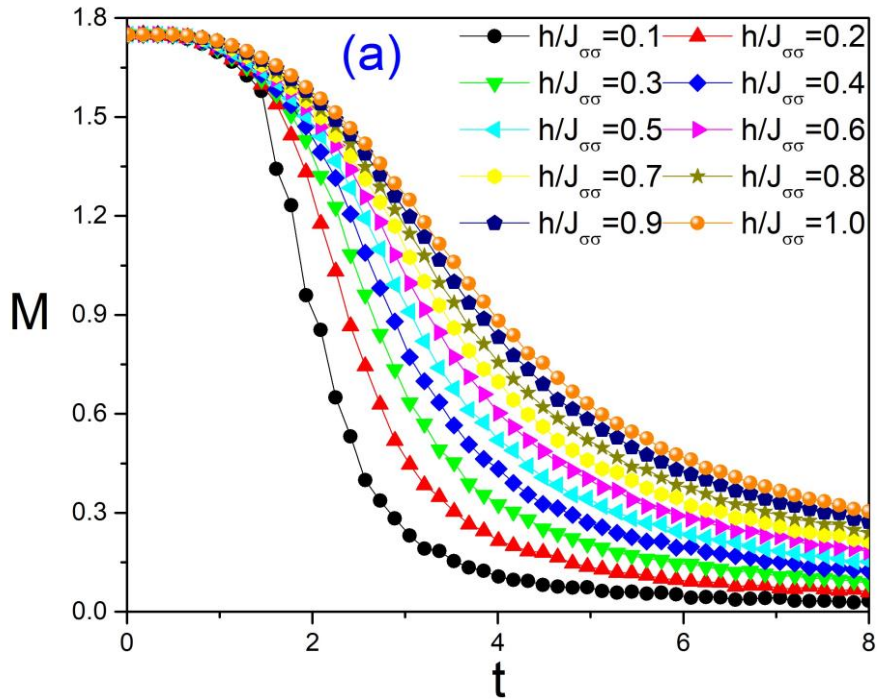

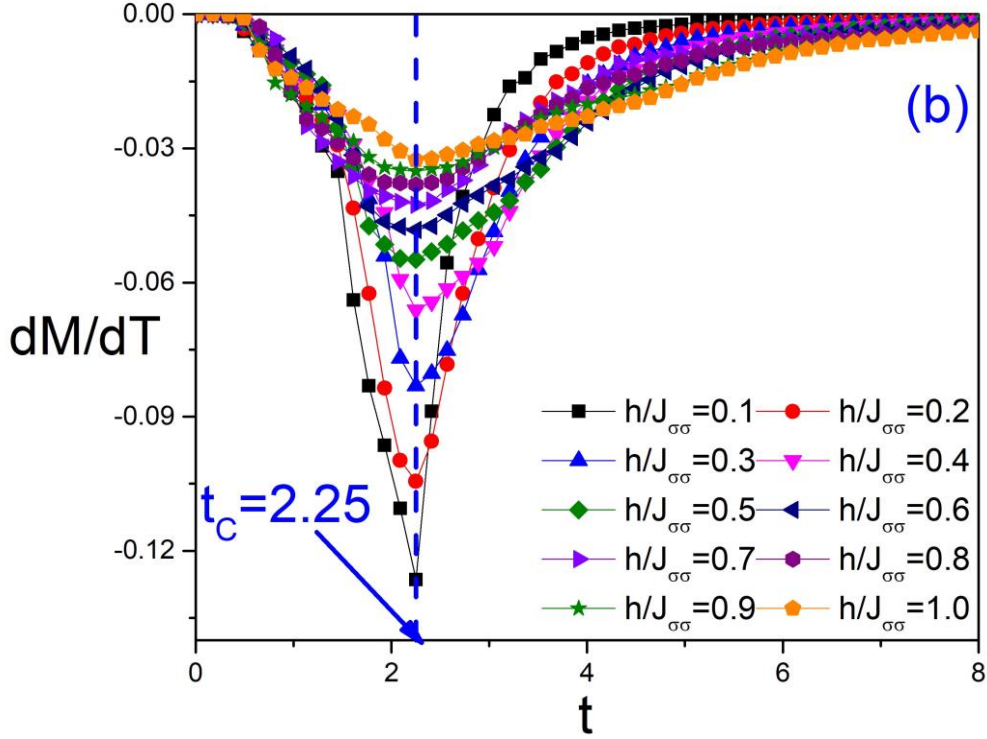

*Fig. 3: The total magnetizations (a) and dM/dT (b) of the $C_{20}$ system with mixed spins S=2 and σ =3/2 for r=0.2, p=1.0 and d=0.*

The focus of this section is on the investigation of the magnetocaloric effect (MCE), which is a characteristic of magnetic materials. In Figs. 4(a) and 4(b), we represent the reduced temperature dependence of $\Delta S_m$ curve at different reduced external magnetic fields $h/J_{\sigma\sigma}$ ranging from 0.1 to 1.0 for p=1.0 and d=0.0, respectively. The negative sign of the $\Delta S_m$, which indicates that heat is released when $h/J_{\sigma\sigma}$ is altered adiabatically, is evident from the figures. The value of $\Delta S_m$ grows up to a maximum value where the reduced temperature approaches $t_C$ and then decreases with increasing reduced temperature. Additionally, by increasing the reduced interactions between the spins S from 0.2 to 1, we observe an increase in the value of $t_C$ from $t_C \approx 2$ to $t_C \approx 3$. An increase of $h/J_{\sigma\sigma}$ enhances $\Delta S_m^{MAX}$ and shifts the $\Delta S_m^{MAX}$ point gradually towards higher reduced temperatures, it suggests that for greater reduced magnetic fields, a significantly larger entropy change should be anticipated (see Fig. 5).

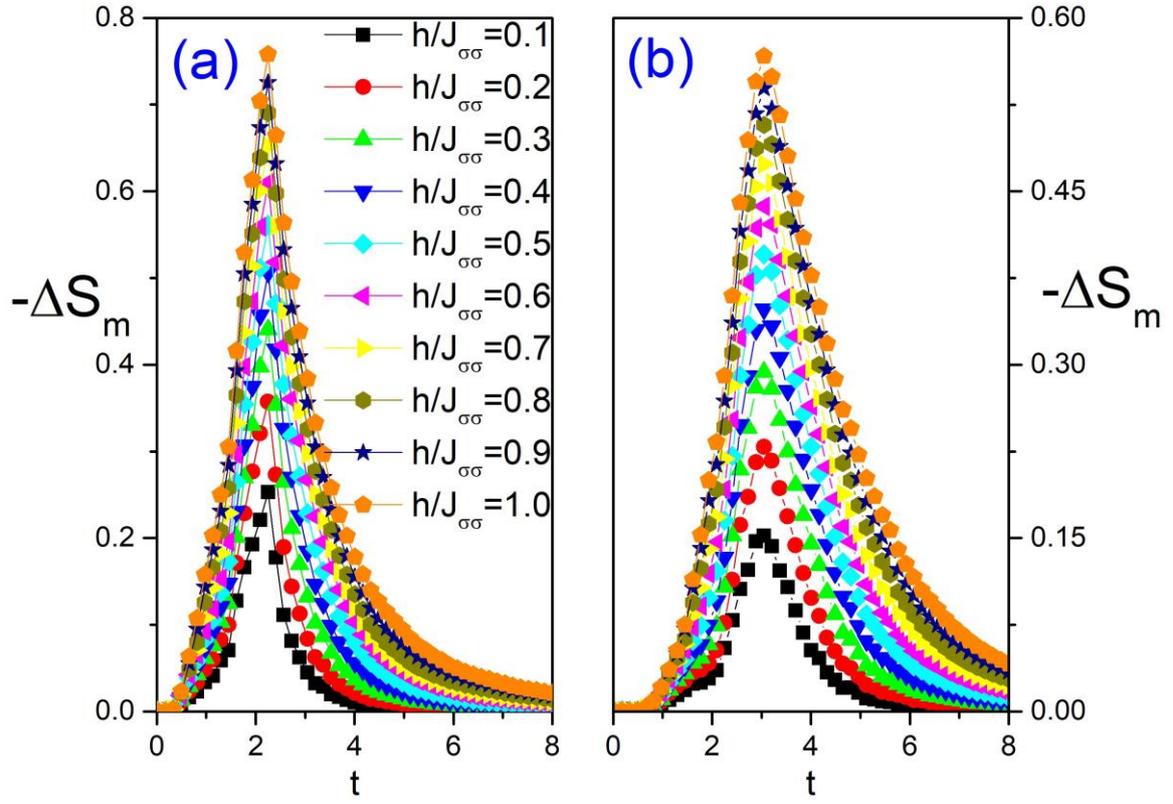

*Fig. 4*: Temperature dependence of the magnetic entropy change of reduced temperatures of $C_{20}$ system for several reduced external magnetic fields $h/J_{\sigma\sigma}$ from 0.1 to 1.0 and r=0.2 (a), r=1.0 (b) for p=1.0 and d=0.0

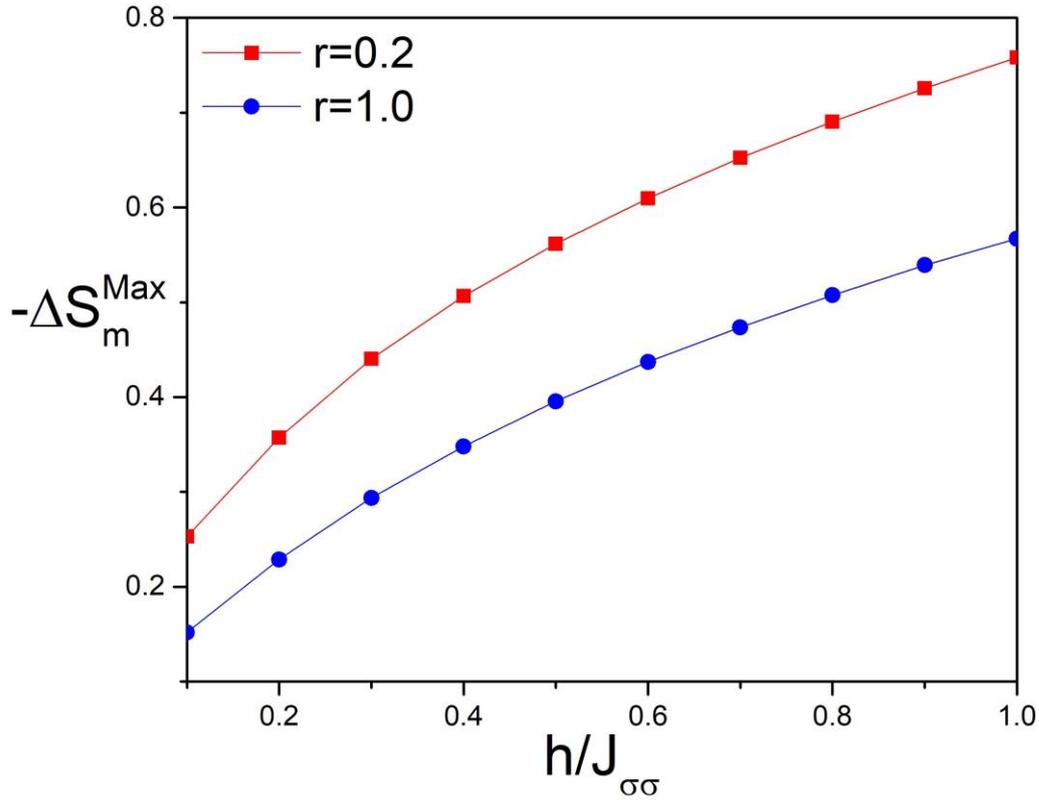

***Fig. 5***: *The maximum magnetic entropy changes of reduced external magnetic fields of $C_{20}$ system for r=0.2 and r=1.0 with p=1.0 and d=0.0.*

The relative cooling power (RCP), a crucial metric, describes the magnetic materials in order to better comprehend the impact of the magnetocaloric features. In an ideal refrigeration operation, the RCP parameter quantifies the quantity of heat that is transported from the heat source to the cold source. Fig.6 presents the relative cooling power versus reduced external magnetic field for r=0.2 and r=1.0 with p=1.0 and d=0. The relative cooling power increases with increasing reduced external magnetic field and RCP increases with increasing r.

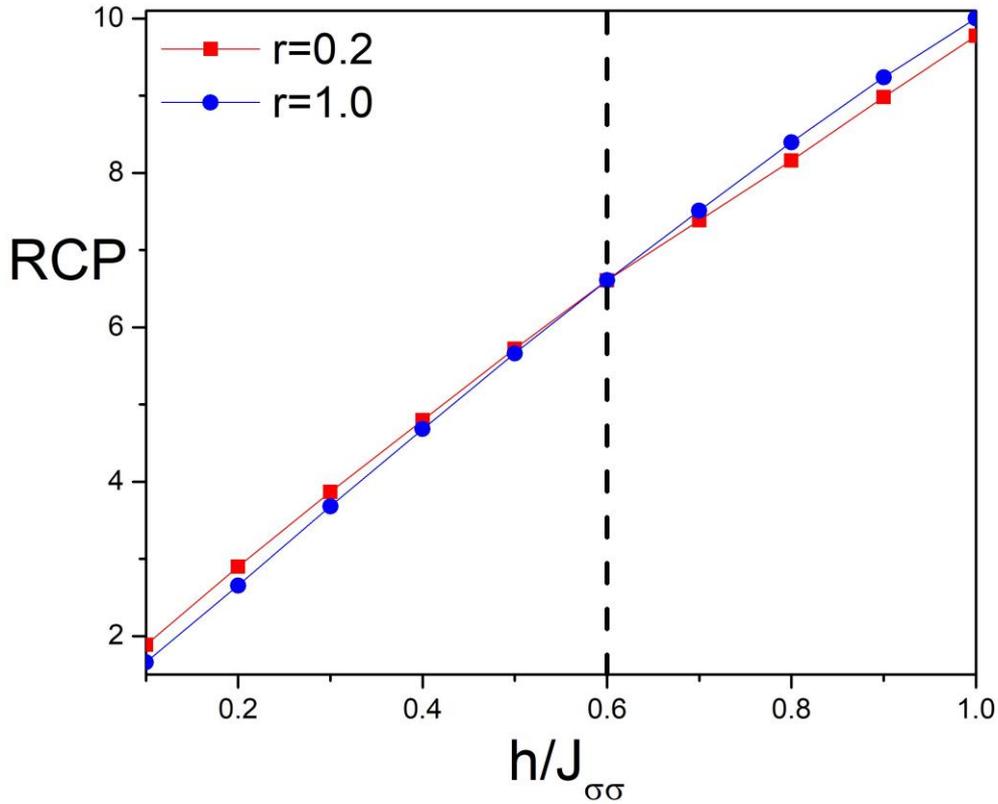

***Fig. 6:*** *The field dependence of relative cooling power (RCP) of $C_{20}$ system for r=0.2 and r=1.0 with p=1.0 and d=0.*

The total magnetization versus the reduced crystal field for reduced temperature t=0.5, 1.0, 1.5 and 1.8 and r=1.0 is shown in Fig. 7(a) with p=1.0 and $h/J_{\sigma\sigma}$=0.1. We can see that the magnetization increases with increasing the value of reduced crystal field until reached its saturation and decrease with increasing t. Moreover, Fig. 7(b) is plotted for r=0.7, 1.0, 1.7 and 2.0 and t=1.0 with p=1.0 and $h/J_{\sigma\sigma}$=0.1. It was shown that the magnetization increases with increasing the value of reduced crystal field and r until reached its saturation.

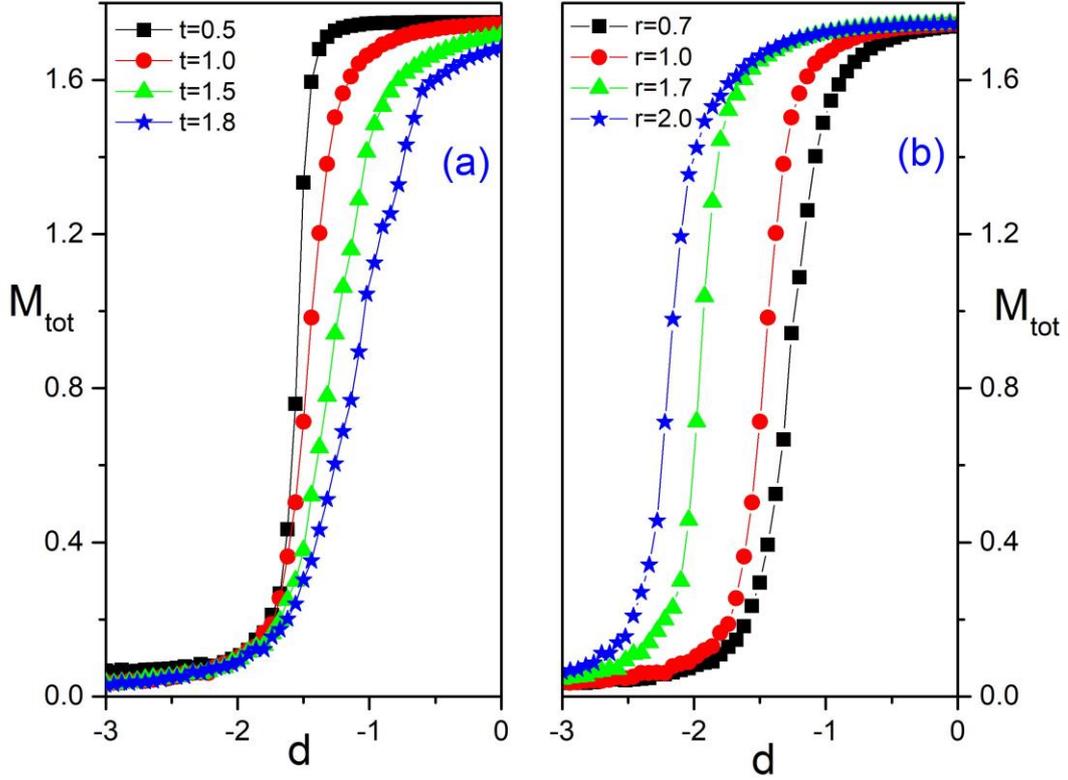

***Fig.7:*** *The total magnetizations versus the reduced crystal field d of **$C_{20}$** system with mixed spins S=2 and σ=3/2 for t=0.5, 1.0, 1.5, 1.8 and r=1.0 (a) and r=0.7, 1.0, 1.7, 2.0 and t=1.0 (b) with p=1.0 and $h/J_{\sigma\sigma}$=0.1.*

Finally, we present in Figs. 8(a) and 8(b), the magnetic hysteresis cycles for: t=0.5, 1.0, 2.5 and 4.0, r=1.0, p=1.0 and d=0.0. In fact Fig. 7(a) corresponds to r=0.2, 1.0, 1.7, 2.8, t=1.0, p=1.0 and d=0.0 in Fig. 7(b), p=0.1, 0.5, 1.0, 1.6, t=1.0, r=1.7 and d=0.0 in Fig. 7(c) and d=0.0, -0.8, -1.4, -2.1, t=1.0, p=1.0 and r=1.7 in Fig. 7(d). The magnetic remanence remains constant with increasing the parameters such as h, t, d, r and p. In Fig. 7(a) and Fig. 7(c), we see that the reduced magnetic coercive field increases with decreasing reduced temperature t and reduced exchange interaction p. In Fig. 7(b) and Fig. 7(d), we observe that the reduced magnetic coercive field decreases with increasing reduced exchange interaction r and reduced crystal field d. It supports the material's ferromagnetic properties by showing the significant contrast between the applied field's negative and positive values below the transition temperature. Above the transition temperature, they have not shown any remanent magnetization or coercive field. This kind of action is consistent with the substance's paramagnetic properties.

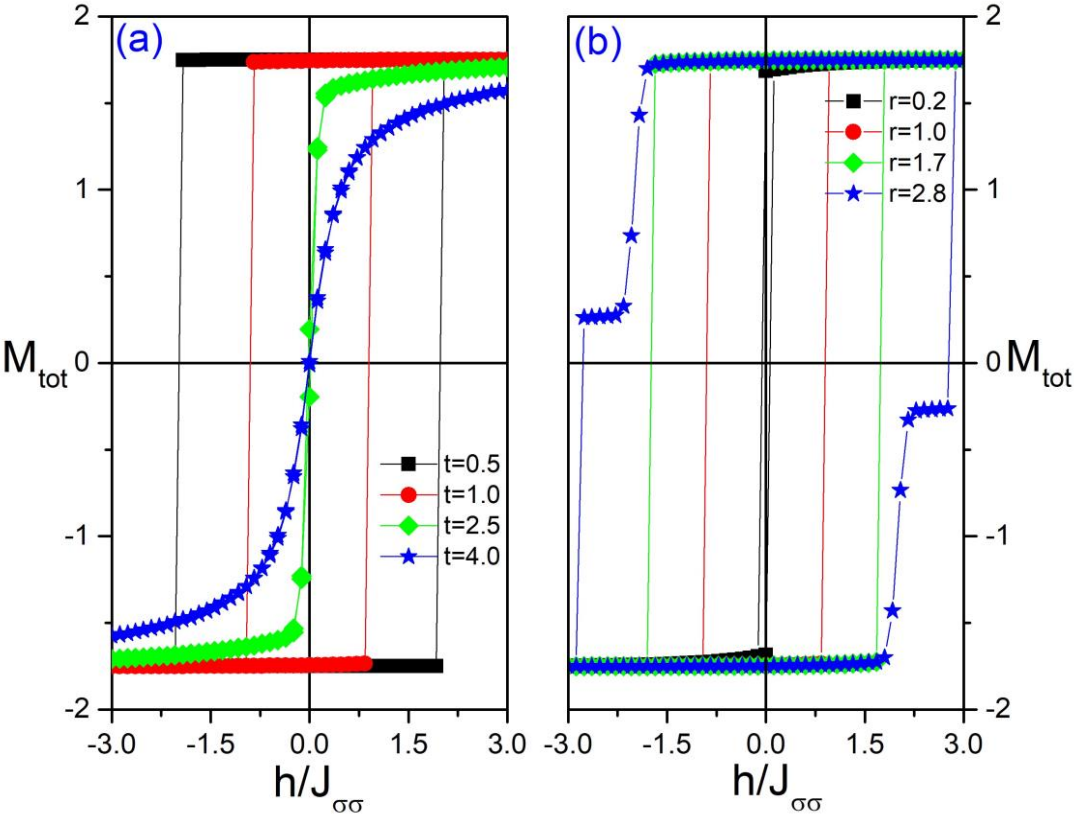

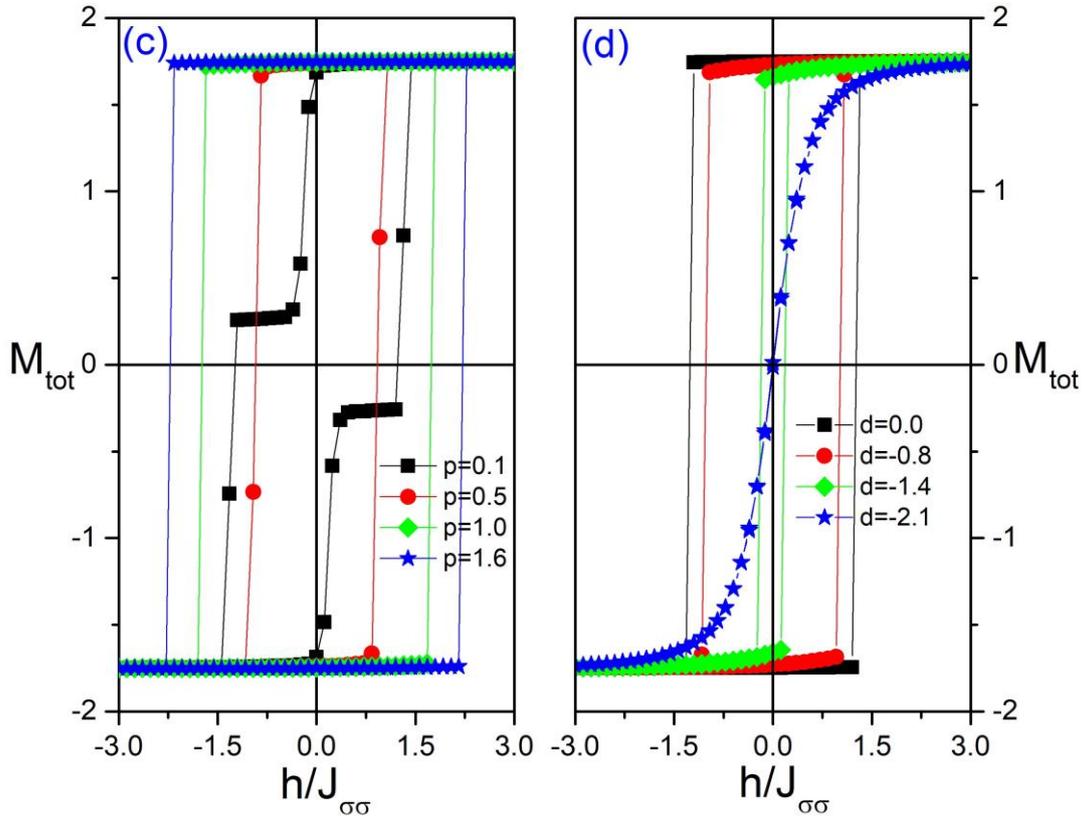

*Fig. 8:* *The magnetic hysteresis cycle of **C₂₀** system for t=0.5, 1.0, 2.5, 4.0, r=1.0, p=1.0 and d=0.0 (a), r=0.2, 1.0, 1.7, 2.8, t=1.0, p=1.0 and d=0.0 (b), p=0.1, 0.5, 1.0, 1.6, t=1.0, r=1.7 and d=0.0 (c) and d=0.0, -0.8, -1.4, -2.1, t=1.0, p=1.0 and r=1.7 (d).*

4. **Conclusions**

In conclusion, Monte Carlo simulations are used to study the magnetic and magnetocaloric properties of the Fullerene **C₂₀** system with mixed spins 2 and 3/2. We show how a reduced crystal magnetic field, a reduced external magnetic field, and the reduced coupling interaction between spins affect the mixed spin system studied. For t=0, the ground state phase diagrams were established in the planes: external magnetic field, coupling interaction between spins and the reduced crystal magnetic field. The behavior of the magnetizations and the derivative of magnetization, in particular, have shown the impact of rising reduced temperature. At a lower critical reduced temperature, the system can display a phase change from the ferromagnetic to the paramagnetic phase. For numerous reduced external magnetic fields and reduced temperature, the magnetic entropy variations are discovered. It is discovered that the reduced exchange coupling interactions p and r lead to an increase in the lowered magnetic reduced coercive field.